# ONTOLOGY SERVICE CENTER: A DATAHUB FOR ONTOLOGY APPLICATION

CHEN Tao[1], SU Rina[2], ZHANG Yongjuan[3], YIN Xin[4] and ZHU Rui[5]

[1]School of Information Management, Sun Yat-sen University, Guangzhou, Guangdong, China.
[2]Library, Sun Yat-sen University, Guangzhou, Guangdong, China.
[3]Shanghai Information Center for Life Sciences, Chinese Academy of Sciences, Shanghai, China.
[4]School of Information Resource Management, Remin University of China, Beijing, China.
[5]Department of Library, Information and Archives, Shanghai University, Shanghai, China.

## ABSTRACT

*With the growth of data-oriented research in humanities, a large number of research datasets have been created and published through web services. However, how to discover, integrate and reuse these distributed heterogeneous research datasets is a challenging task. Ontology is the soul between series digital humanities resources, which provides a good way for people to discover and understand these datasets. With the release of more and more linked open data and knowledge bases, a large number of ontologies have been produced at the same time. These ontologies have different publishing formats, consumption patterns, and interactions ways, which are not conductive to the user's understanding of the datasets and the reuse of the ontologies. The Ontology Service Center platform consists of Ontology Query Center and Ontology Validation Center, mainly using linked data and ontology-based technologies. The Ontology Query Center realizes the functions of ontology publishing, querying, data interaction and online browsing, while the Ontology Validation Center can verify the status of using certain ontologies in the linked datasets. The empirical part of the paper uses the Confucius portrait as an example of how OSC can be used in the semantic annotation of images. In a word, the purpose of this paper is to construct the applied ecology of ontology to promote the development of knowledge graphs and the spread of ontology.*

## 1. INTRODUCTION

With the growth of data-oriented research in humanities, a large number of research datasets have been created and published through web services. A large part of these datasets is published on web as linked open data. As we all know, the purpose of Linked Data is to aggregate, harmonize, integrate, enrich and publish data for reuse on the Web in a cost-efficient way when using Semantic Web technologies (Bizer et al., 2009; Tom et al., 2011; Gandon, 2018). However, how to discover, integrate and reuse these distributed heterogeneous research datasets is a great challenge. As ontology becomes the logical backbone for managing research datasets in the humanities, developing ontology repositories to archive semantic interoperability among research datasets becomes a viable solution to address this issue. Although many datasets publish their corresponding ontologies, the others don't publish their ontologies simultaneously. The lack of ontology description creates a comprehension barrier of the datasets use, also consumers cannot clearly understand the data structure of the datasets. Finland scholars have proposed the extending of the 5-Star Model proposed by Tim Berners-Lee in 2010 into a 7-Star Model, intending to encourage data publishers to provide their data with explicit metadata schemas (Hyvönen et al., 2014).

There are many studies on ontology, which can be divided into the following research spots. First, there are a large number of articles on how to design and construct ontologies and knowledge graphs. Wang Xiaoguang et al. (2020) construct an ontology model to regulate the entities, attributes and relationships of Dunhuang cultural heritage knowledge. In recent years, Shanghai Library has published several digital humanities-related ontologies (Xia C.J. et al., 2014; 2021). Muir K. et al. (2018) discuss ontology-based information retrieval approaches and techniques by taking into consideration the aspects of ontology modeling, processing and the translation of ontological knowledge into database search requests.

Second, there are also many scholars studying ontological specifications and recommendations to ensure the efficient publishing of ontology, such as the OBO Foundry (Smith B. et al., 2007) establishes a set of principles for ontology development for creating a suite of interoperable reference ontologies in the biomedical domain. Chen T. et al. (2019) and Feitosa, D. et al. (2018) give several specifications and recommendations on Linked Data publishing. Garijo D. et al. (2020) describes guidelines and best practices for creating accessible, understandable, and reusable ontologies on the Web.

Besides, there're studies are focusing on how to build a unified platform to manage and maintain multiple ontologies in the bio-medical field. BioPortal (Whetzel, P.L. et al., 2009; Noy, N.-F. et al., 2011; Vescovo, C.D. et al., 2011) is an open-source online community-based ontology repository that has been used as a critical component of semantic infrastructure in several domains, including biomedicine and bio-geochemical. By reusing the National Center for Biomedical Ontologies (NCBO) BioPortal technology, Jonquet C. et al. (2018) have designed AgroPortal, a vocabulary and ontology repository for agronomy, food, plant sciences, and biodiversity. Codescu M. et al. (2017) proposed Ontohub, a semantic repository for heterogeneous ontologies. Linked Open Vocabularies (LOV) (Vandenbussche, P.-Y. et al., 2017) is a high-quality catalogue of reusable vocabularies for the description of data on the Web that aims to promote and facilitate the reuse of well-documented vocabularies in the Linked Data Ecosystem.

The above analysis mainly revolves around the ontology design principles, design methods and maintenance management, and in this paper, we will focus on the application of ontologies. As everyone knows, a larger number ontologies are used for digital humanities projects in the way of knowledge organization, some of which are based on common ontologies, many of which are designed according to project requirements. After design ontologies, most of these ontologies are still stored separately in different structures and datasets published by dataset providers, which will lead to a variety of data formats and consumption methods for the published ontologies. In terms of data formats, some are in RDF/XML formats while others are in TTL, N3, or JSON-LD in terms of acquisition methods, some provides ontologies online browse, and can only be downloaded via the static ontology files. Therefore, we design and implement the Ontology Service Center (OSC) platform, which belongs to the third category of ontology research discussed above.

In general, the OSC platform has many same functions with LOV platform, such as ontology or vocabulary registry, publishing, querying and download. However, OSC platform provide some advanced and more convenient functions and services:

1) In addition to the retrieval of ontology and properties, the OSC platform also provides a variety of online browsing methods for ontology, while the ontology file needs to be downloaded in LOV.
2) OSC has more diverse and richer ontology interaction formats and these different serialization formats will be converted automatically as requested and the ontology format in LOV is mainly N3.
3) LOV is mainly used for ontology searching, OSC can also be used for dataset verification, which is mainly used to verify the usage status of ontologies in the dataset.

The rest of this article is organized as follows. In the next section, we describe the framework of the OSC platform and the storage mechanism of ontology. In Section 3, we introduce the ontology query service module and explain the principle and algorithm of property inheritance. Moreover, we describe how to validate instance data with ontology in Section 4. In Section 5, we illustrate how to use OSC for image semantic annotation in conjunction with the IIIF-IIP platform.

## 2. SYSTEM COMPONENTS AND FEATURES

### 2.1. Ontology Service Center (OSC) Framework

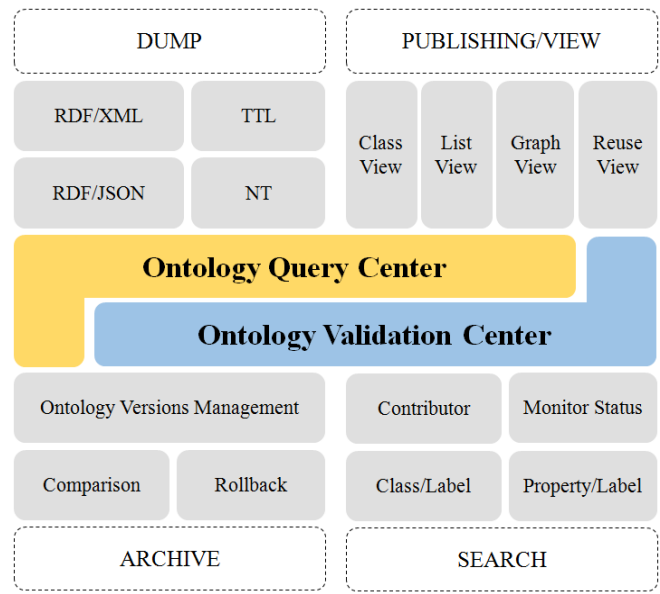

Figure 1. Ontology Service Center Framework

The Ontology Service Center (OSC) Framework, shown in figure 1, is composed of four main components: data dump, data publishing/view, data archive and data search. Based on these features, the system provides two major ontology services, Ontology Query Center (OQC) and Ontology Validation Center (OVC). OQC focuses on the ontology management process, which is used to store and publish ontologies, while OVC validate ontology, which is used to verify the use of ontologies in linked open data. It is important to note that OVC does not perform syntax checks about ontology. When ontology has a syntax error, it cannot be registered in OQC.

**Data Dump.** OSC provides four data dumps, RDF/XML, TTL, RDF/JSON and NT, which will be generated automatically in real-time of ontology (RDF) model which is a set of Statements of this ontology.

**Data Publishing/View.** This system supports four kinds of ontology view methods, Class View (C.V.), List View (L.V.), Graph View (G.V.) and Reuse View (R.V.).

- Class View. This view uses a tree structure to display the ontology, which makes it easy to understand the hierarchical relationship between classes in ontology.
- List View. This view presents the entire vocabulary on one page with an ordered list of classes and properties, followed by more detailed information panels further down the document.
- Graph View. This view visually displays the ontology with WebVOWL (Lohmann, S. et al., 2015; Wiens, V. et al., 2018), which is a web application for the interactive visualization of ontologies. The visualization is automatically generated from the ontology graph.
- Reuse View. This view is achieved through the WebVOWL Editor application, which is designed to serve the skills and needs of domain experts with limited knowledge of ontology modeling.

**Data Archive.** The archived ontology is served as a file on disk, and the file name will contain the version number of the archive. In other words, only the latest ontology will be stored in RDF format. The system will provide the results of the comparison between the archived version and the latest one. If necessary, the archived ontology can be rolled back and restored to the latest version in the ontology graph.

**Data Search.** OSC enables searching for vocabulary terms (class, object property, data property), term comments, catalogues, namespace prefix and contributor.

## 2.2. Ontology Storage Mechanism Design

In the OSC system, we use OpenLink Virtuoso (Triple Store) to store the latest ontologies. Figure 2 shows the storage mechanism of this system. Every ontology with the latest version is stored independently in a named graph. That is, as many ontologies, there are so many named graphs in the triple store. At the same time, we need to establish a management graph of datasets called datasets graph here, to manage the basic information of ontologies, such as ontology publisher, publishing time, version number, rights and license. This dataset graph is the core hub of the entire ontology management mechanism which is used to manage all ontologies and versions.

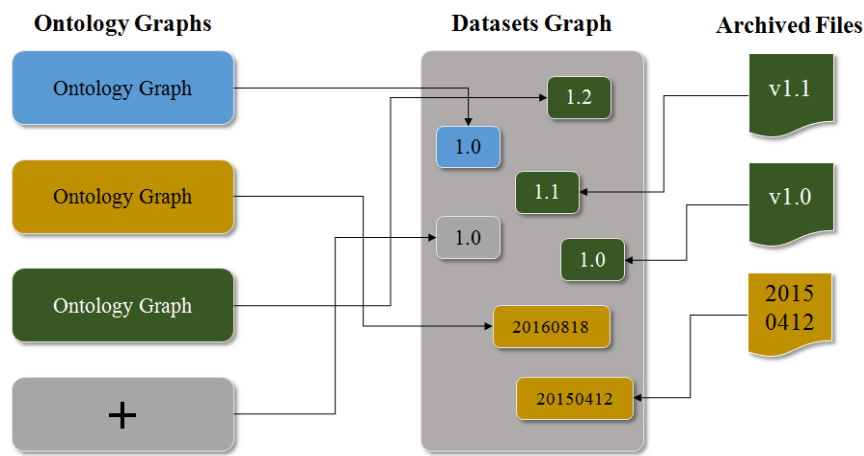

Figure 2. Ontology Storage Mechanism

Next, we take an example of Schema.org ontology to explain how to manage ontology and its version files in the triple store. Schema.org provides a collection of shared vocabularies webmasters can use to mark up their pages in ways that can be understood by major search engines: Google, Microsoft, Yandex, and Yahoo!. Here, we store Schema.org ontology in two latest versions, v12 and v3.1, in which v12 is the latest version and v3.1 is an archived version. The description of the ontology's metadata, we refer to recommendations by Vandenbussche (Vandenbussche P.Y., 2012).

**Example 1-a: Schema.org ontology in 12 version (latest version)**

<http://ont.library.sh.cn/ontology/schema_12.0> a owl:Ontology ;
dc:title "Schema.org vocabulary (schema)" ;
   dc:description "Search engines including Bing, Google, Yahoo! and Yandex rely on schema.org markup to improve the display of search results, making it easier for people to find the right web pages." ;
   owl:versionInfo "12.0" ;
dc:rights "© 2021 Schema";
   dc:issued "2021-03-08 00:00:00" ;
cc:license <https://createivecommons.org/licenses/by/3.0/> ;
 **dc:source** <http://ont.library.sh.cn/graph/schema>.

**Example 1-b: Schema.org ontology in 3.1 version (archived version)**

<http://ont.library.sh.cn/ontology/schema_3.1> a owl:Ontology ;
dc:title "Schema.org vocabulary (schema)" ;
   dc:description "Search engines including Bing, Google, Yahoo! and Yandex rely on schema.org markup to improve the display of search results, making it easier for people to find the right web pages." ;
   owl:versionInfo "3.1" ;
dc:rights "© 2016 Schema";
   dc:issued "2016-08-09 00:00:00" ;
cc:license <https://createivecommons.org/licenses/by/3.0/>.

When ontology has a new version, the previously stored in named graph one will be archived. In this example, we can find two main differences. One is that the resource URIs of these two versions are different. We add the version number in the resource URI of the ontology version, which ensures that different ontologies and versions are strictly distinguished. The other difference is that there is dc:source property in the latest version of Schema.org ontology. This property plays a very important role in ontology management and the value of this property is the named graph that stores the latest version of this ontology. When retrieving Schema.org ontology, the OSC system will quickly find the latest version of this ontology in multiple named graphs.

Similarly, when you want to obtain the historical versions of Schema.org, you can query resources that don't contain dc:source property in the datasets graph of this ontology. The historical versions are then found through the version number property (owl:versionInfo) or the time property (dc:issued). With the version number, it is very easy to find the required ontology version from archived files.

## 3. ONTOLOGY QUERY CENTER/SERVICE (OQC)

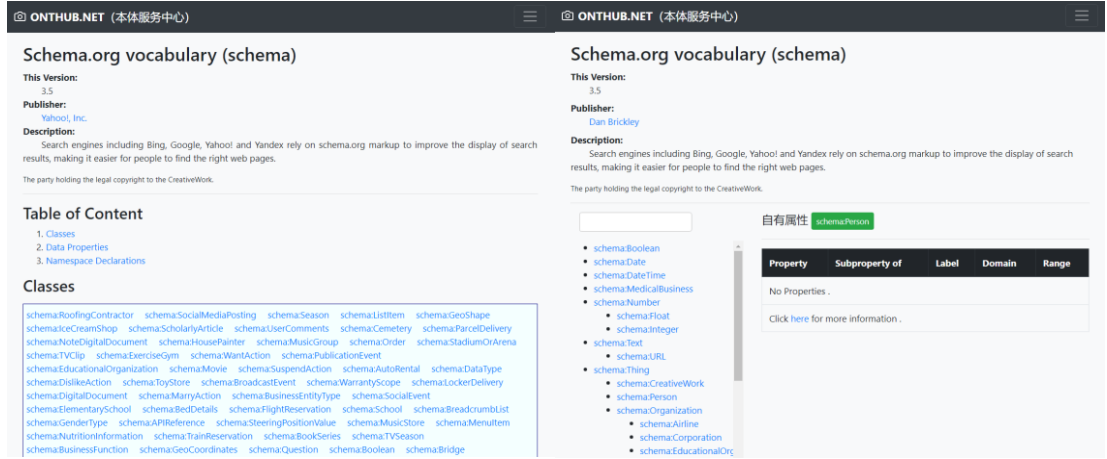

(a) List View          (b) Class View

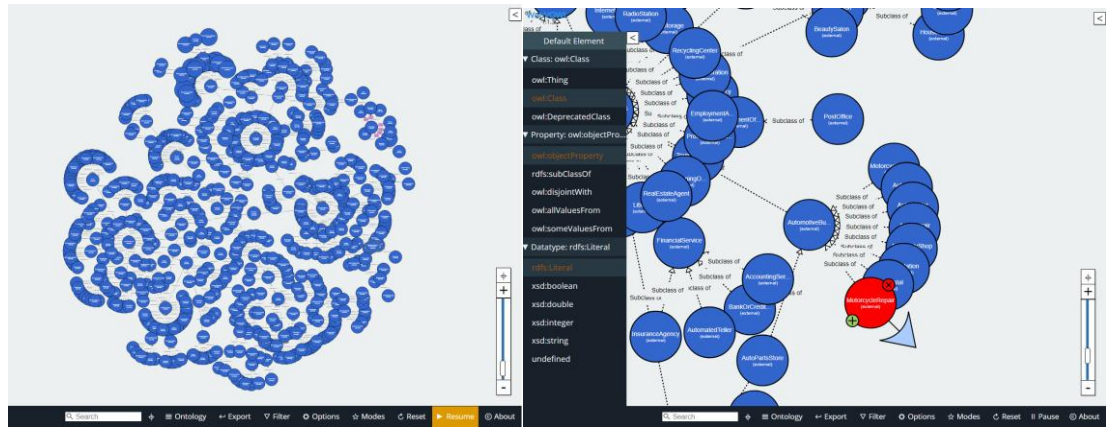

(c) Graph View          (d) Reuse View

Figure 3. Schema.org Ontology in Onthub Center

Ontology Query Center provides the function of acquiring and viewing ontology. Figure 3 shows the function snapshots of Schema.org ontology in the OSC platform. Figure 3-A to 3-D are the different ways to browse this ontology. 3-A is the list view (L.V) mode, and 3-B is the class view (C.V) mode. 3-C and 3-D modes are performed with the WEBVOWL plug-in. Class view mode not only shows the hierarchical relationship between different classes of ontology, but also shows the inheritance relationships between properties. Next, we will introduce the property path calculation algorithm. Figure 4 is the diagram of property inheritance and assuming that each layer class contains its private properties in addition to the properties of the previous classes. At the same time, the class can also inherit from multiple parent classes.

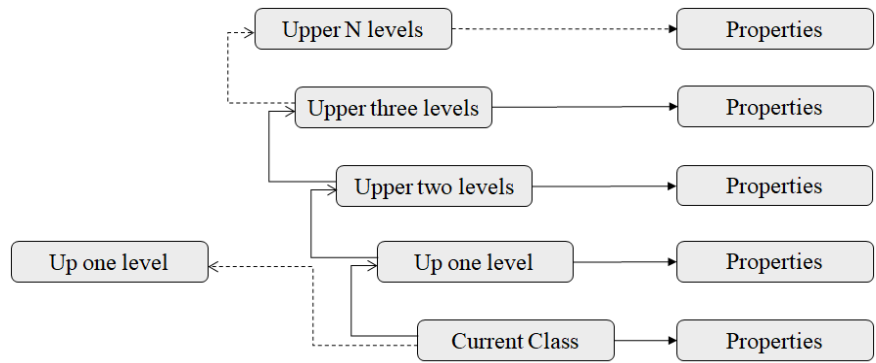

Figure 4. Diagram of Property Inheritance

During implementation, we need to find the top-level classes of the ontology. The top-level class means that the class is only inherited by other subclasses, and doesn't inherit other classes. With these top-level classes, the inherited classes will be found in turn. And the SPARQL of the core part is as follows:

```
//Get top-level classes
SELECT DISTINCT ?class ?label
FROM <graph_uri>
WHERE {
        VALUES ?type {owl:Class rdfs:Class}
        {
                ?class a ?type .
                FILTER (isIRI(?class))
        } UNION {
                ?a a ?type ; rdfs:subClassOf* ?class .
                FILTER (isIRI(?a)) FILTER (isIRI(?class))
        }
        OPTIONAL {?class rdfs:label ?label .}
        FILTER NOT EXISTS{?class rdfs:subClassOf ?sub}
} ORDER BY ASC(STR(?class))

//Get subclasses, which need to execute recursively in programming
SELECT DISTINCT ?class ?label
FROM <graph_uri>
WHERE {
        ?class rdfs:subClassOf <class_uri> .
        OPTIONAL {?class rdfs:label ?label . }
}
```

The SPARQL in the first part explains how to get the top-level classes of all non-blank nodes in the ontology graph. The types of top-level classes here are mainly owl:Class and rdfs:Class. The second SPARQL is used to obtain the inherited classes of a certain class (<class_uri> should be replaced by the special class URI). It should be noted that when the algorithm is implemented, the query statement needs to be executed recursively until the end. After obtaining all the classes and structure levels, you can use some javascript libraries to display the ontology tree in the site. The following example of "shlnames: Shanghai Library Names Authority Ontology" illustrates the practical application of property inheritance shown in figure 5. The current class is shl:Person which is the subclass of shl:Agent and foaf:Person and the foaf:Person is also the subclass of rdfs:Resource. As can be seen in this figure, we have defined many properties and only rel:childOf, rel:friendOf and rel:influenceBy are listed here. In the property inheritance section, we see that there are two levels here, shl:Agent (up one level) and rdfs:Resource (upper two levels). The foaf:Person class has no relevant properties, therefore, it is not rendered in this section. As accessing the related inherited parent class, you can see the related properties.

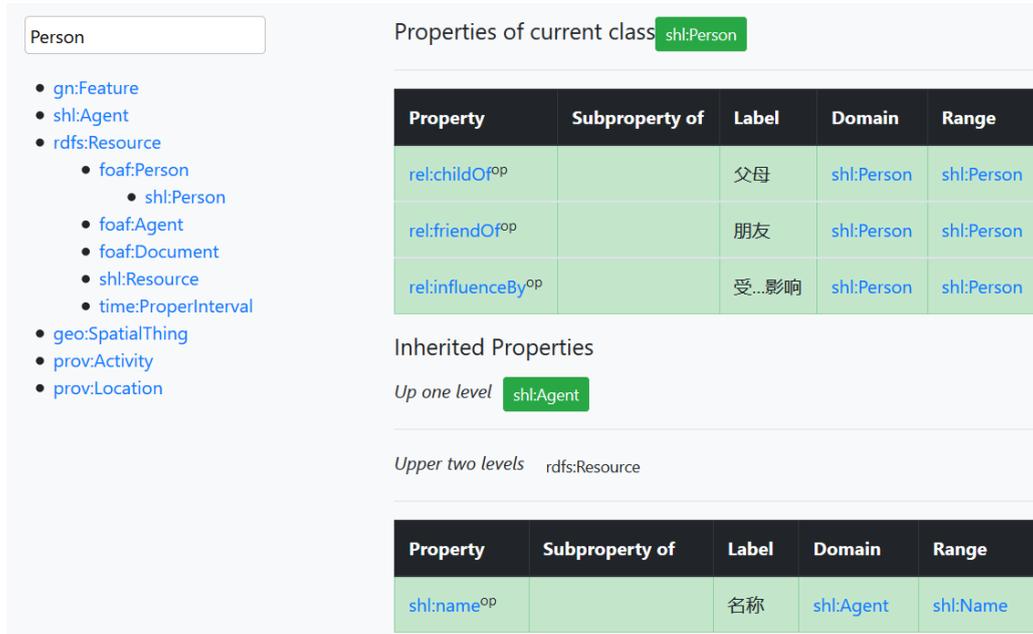

Figure 5. shlnames Ontology in Tree View

## 4. ONTOLOGY VALIDATION CENTER/SERVICE (OVC)

Using ontology validation centers can solve two common types of scenarios. One is ontology validation, aiming to validate the classes and properties in RDF data according to the published ontology file. Data structures in the RDF data published by the dataset often change as data increases. However, the published ontology is not synchronized due to a failure to update it promptly. Data consumers are confused when they access these RDF data because there is no description of these newly added or discarded properties in the published ontology. The other scenario is ontology statistics, using a validation service to count the number of instances corresponding to the classes and properties in the ontology. With statistical information, it is convenient for consumers to have a holistic understanding of the dataset and to make a correct assessment of it.

The ontology validation center contains two parameters, "SPARQL endpoint" and "ontology". The ontology URI should be selected from the ontology query center. When both parameters have input values, the ontology validation process is started. If only enter the SPARQL endpoint and don't enter the ontology URI, ontology statistics are performed. Now, we will use an example to illustrate this transaction logic. The China Biographical Database (CBDB) is a free accessible relational database with biographical information about approximately 422,600 individuals, primarily from the 7th through 19th centuries. In the ontology query center, we have created CBDB ontology which is used in CBDB Linked Data (CBDB_LD) dataset, and the ontology URI is "http://ont.library.sh.cn/graph/cbdb". CBDB_LD provides a SPARQL endpoint with the address, "http://cbdb.library.sh.cn/sparql". It is also important to note that all graphs for multiple projects are often stored in the same RDF Store. Therefore, we need to set the target graphs before executing validations or statistics.

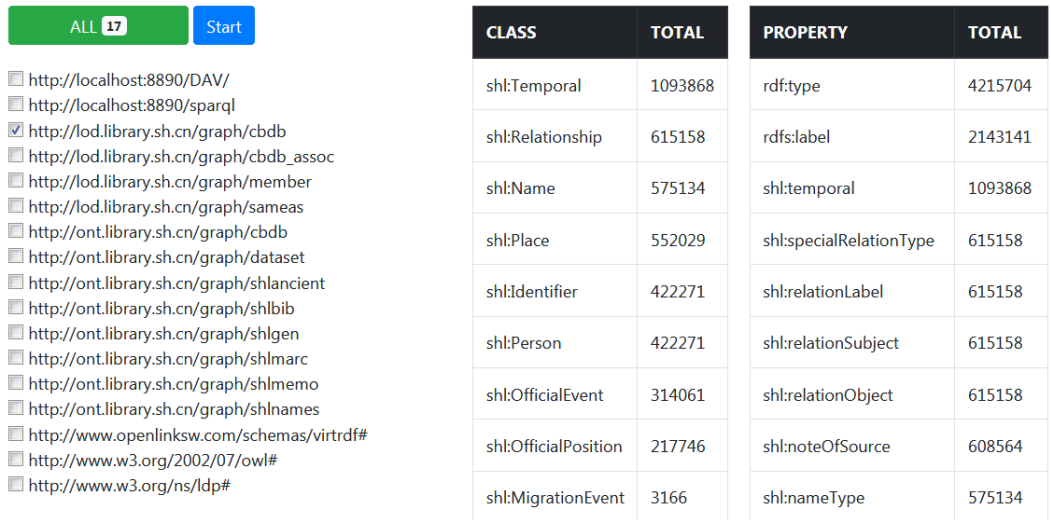

Figure 6. Ontology Statistics for CBDB_LD Dataset in OVC

Figure 6 shows the statistical results of CBDB_LD, where we only entered the endpoint value. When the "Query Graph" button is executed, all graphs on the left side of Figure 4 are shown, and there are 17 graphs in this dataset. We choose the target graph "http://lod.library.sh.cn/graph/cbdb" and execute "Start". All the classes and properties used in this dataset are listed in tables. It is clear to see that this dataset involves nine classes, just as: shl:Temporal, shl:Relationship, shl:Name, shl:Place, et al. Due to the excessive number of the properties, we have intercepted the nine most used properties in the property statistics table. From this table, the property which use the most is rdf:type, followed by rd fs:label and shl:temporal.

As we enter the ontology URI and perform the same steps, a different result will come out shown in figure 7. There is an icon about the variable result after each class and property record in tables. In the table of the class validation, we can see that all classes are already defined in the ontology which we want to validate. While in the property checking table, there are two properties, shl:relationObject and shl:nameType, marked with failure icon which means that these are not defined in the ontology file. With these validation tables, it is easy to find out whether the published ontology is synchronized with the RDF dataset.

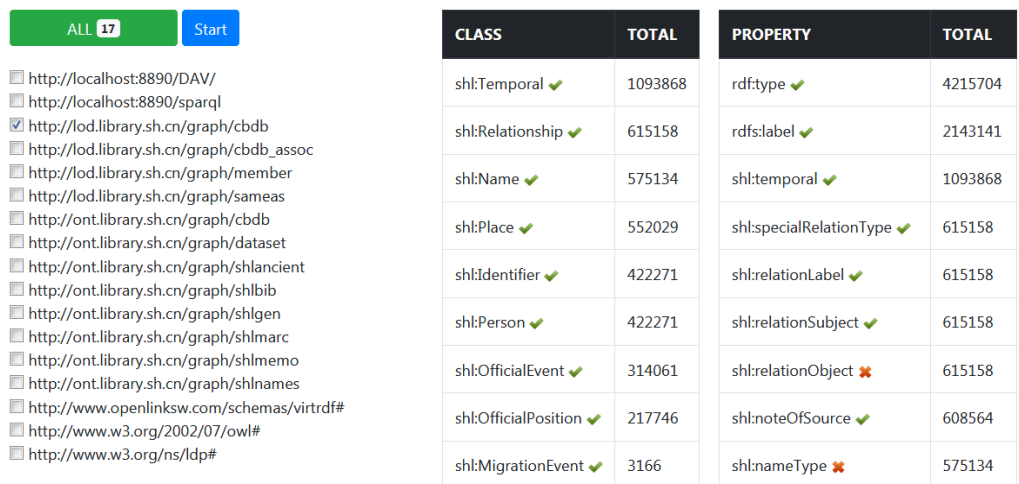

Figure 7. Ontology Validation for CBDB_LD Dataset in OVC

The following describes the core algorithm process of the validation module, which mainly relates to how to manipulate triple data in the ontology graph and the target dataset graph shown in figure 8. These steps are not difficult to implement, therefore, we don't go into too much detail here.

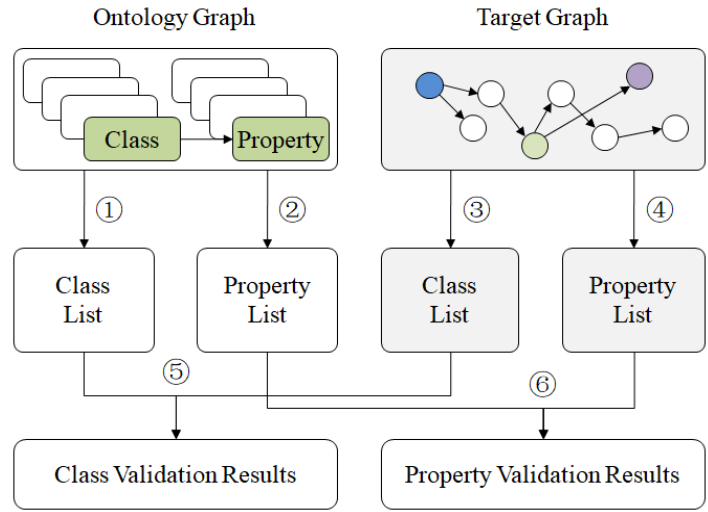

Figure 8. Validation algorithm process

(1) All classes and properties are extracted from the ontology graph, mainly in step 1 and 2.
(2) Steps 3 and 4 are used to extract the classes and properties used in the instance data in the target dataset graph.
(3) In steps 5 and 6, the results compare the previously extracted classes and properties to form comparison results.

## 5. OSC IN IMAGE SEMANTIC ANNOTATION

The OSC platform currently contains hundreds of ontologies from more than 50 domains and has been used by many scholars. Most of these ontologies come from LOV, others from our actual digital humanities programs and the results of other Chinese researchers. In this section, we will illustrate how to use OSC for semantic annotation of images in conjunction with the IIIF-IIP platform . IIIF-IIP is a multi-dimensional image smart system based on IIIF (International Image Interoperability Framework) and this platform mainly uses Image API 2.1, Presentation API 2.1, Search API 1.0, and Web Annotation Data Model, as well as some linked data technologies. It provides a convenient IIIF resources publishing process and lowers the technical threshold for the release and reuse of IIIF resources. IIIF-IIP supports the online publishing of image resources in multiple formats such as JPG, PNG, TIF, GIF, JPEG, etc. It also supports text recognition, image re-aggregate, and semantic annotation. IIIF-IIP can carry out online interaction of images on a large scale to provide a solid technological foundation for the reuse of cultural heritage. Using the "Confucius Disciples Image" as an example, we first need to mark the edges of the image where Confucius is, which is shown in figure 9.

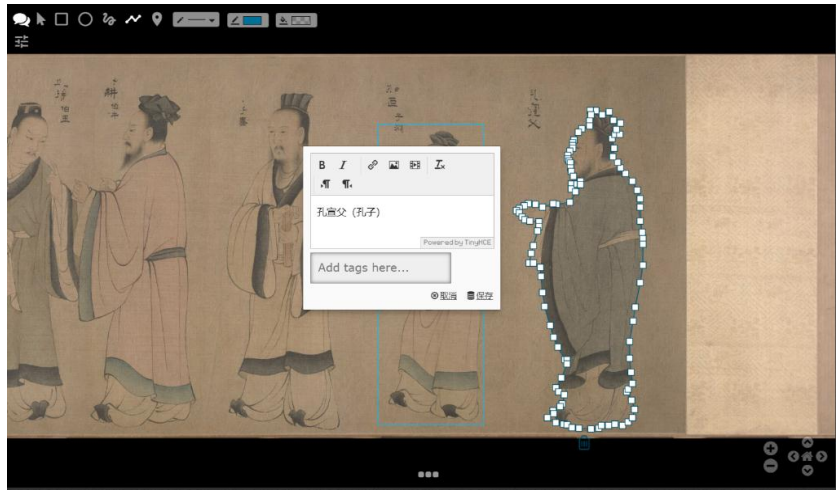

Figure 9. Mark the Edges of Confucius

After tagging the target object in the image, we can connect it to linked open data. When correlated, the OSC platform can be used to query ontologies and properties. owl:sameAs are often used to associate the same resources in different datasets. Other properties can also be used to connect the same resources, such as "related", so we need to find out which properties contain "related" words. After the search, there are more than 80 ontologies in OSC that have the "related" in their properties. Further, we choose Schema.org and skos ontology with properties schema:relatedTo (DatatypeProperty) and skos:related (ObjectProperty). Table 1 shows the data structure of semantic annotation for Confucius with DBpedia knowledge base.

Table. 1. Semantic Annotation in IIIF-IIP

| Key | Value |
|---|---|
| **Property** | http://www.w3.org/2004/02/skos/core#related (skos:related) |
| **Object** | http://dbpedia.org/resource/Confucius |
| **Label** | 孔子(Confucius) |
| **Source** | DBpedia |

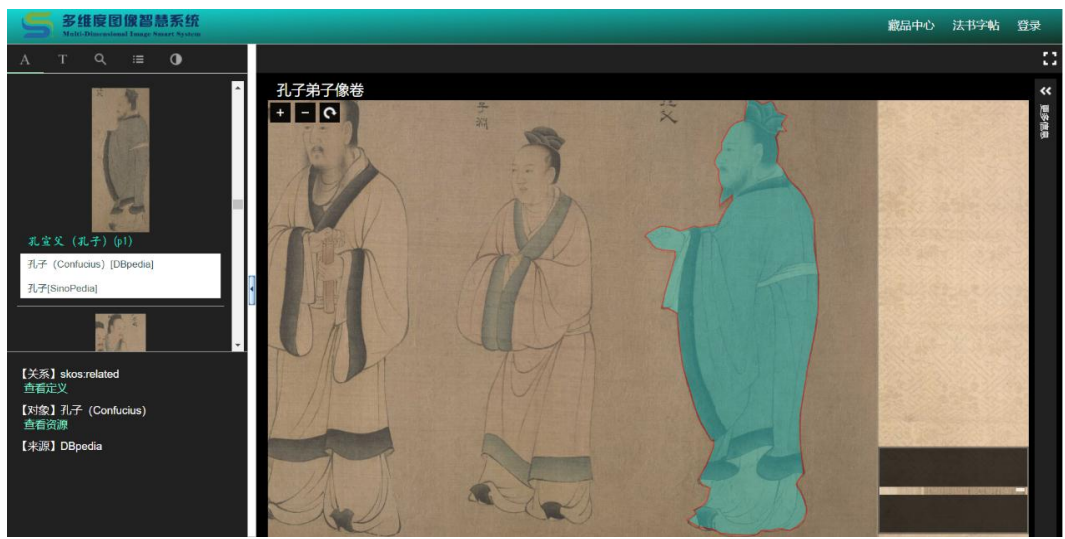

Figure 10. Connect Confucius to DBpedia and SinoPedia knowledge bases

In figure 10, we see that Confucius is precisely marked and when viewing the image, the associated information to linked open data was displayed in left detail panel. You can see more information about Confucius when viewing the connected resources. The knowledge graph of Confucius associated with SinoPedia and DBpedia is shown in figure 11, where the main interests and subjects of Confucius can be easily seen in the graph. The top left node in the figure is the resource from SinoPedia (Chen T. et al., 2019) and the middle node is DBpedia's resource.

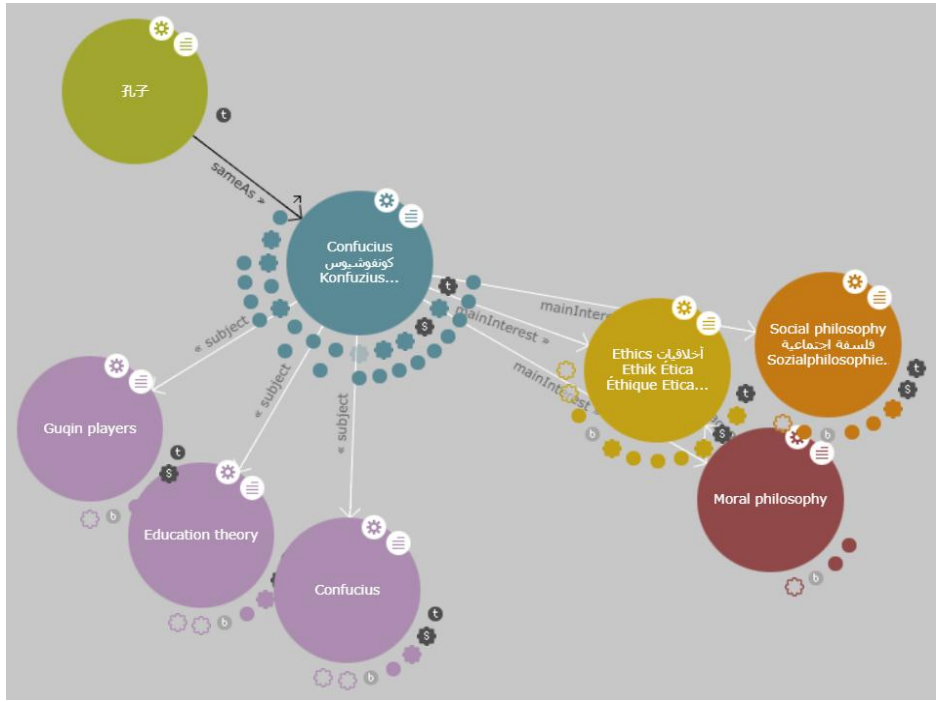

Figure 11. Confucius's Knowledge Graph

## 6. Conclusions

In this paper, we describe Ontology Service Center (OSC) platform which mainly includes Ontology Query Center (OQC) and Ontology Validation Center (OVC) modules. OQC is mainly used for ontology registry, querying, exporting, editing, and version control, while OVC can be used for ontology state verification in RDF datasets. Currently, OSC has been successfully applied in many digital humanities projects in Shanghai Library, which has also been used by other scholars for ontology publishing and reusing. Similarly, OSC has been used in image semantic annotation in IIIF-IIP platform to establish a semantic connection between image contents and the external open datasets.

In addition, some optimization measures in the OSC platform will be considered. First, we will continue to track ontology resources published on the web, and regularly update ontologies on OSC. Secondly, the platform function continues to improve, the automatic comparison and rollback function between different versions of the ontology will be realized. Finally, the user experience will be enhanced, and the platform will provide more convenient interfaces to query and use ontology resources.

## ACKNOWLEDGMENTS

This research is granted financial support from National Social Science Fund of China (19BTQ024).